\documentclass[
reprint,
superscriptaddress,
amsmath,amssymb,
aps,
prb,
]{revtex4-2}
\usepackage[utf8]{inputenc}
\usepackage[english]{babel}
\usepackage[plainpages = false, pdfpagelabels, 
                 bookmarks,
                 bookmarksopen = true,
                 bookmarksnumbered = true,
                 breaklinks = true,
                 linktocpage,
                 colorlinks = true,
                 linkcolor = blue,
                 urlcolor  = blue,
                 citecolor = blue,
                 anchorcolor = green,
                 hyperindex = true,
                 hyperfigures
                 ]{hyperref} 
\usepackage{enumerate}
\usepackage{tabularx}
\usepackage{makecell, multirow}
\usepackage{tikz}
\usetikzlibrary{patterns}
\usepackage{slashed}
\usepackage{physics}
\usepackage{verbatim}
\usepackage{svg}
\usepackage{mathtools}
\usepackage{cancel}
\usepackage{bbold}
\usepackage{bm}
\usepackage{booktabs}
\usepackage[caption=false,singlelinecheck=off]{subfig}
\usepackage{blindtext}
\usepackage{bbm}
\usepackage{graphicx}
\usepackage{dcolumn}
\usepackage{bm}
\usepackage[normalem]{ulem}
\usepackage{xcolor}
\renewcommand{\vec}[1]{\bm{#1}}
\usepackage[percent]{overpic}
\usepackage{subfig}
\newcommand{\moire}{moir{\'e} }
\usepackage{bm}
\renewcommand{\vec}[1]{\bm{#1}}

\begin{document}
\title{Emerging network model in a twisted monolayer--rhombohedral graphene}

\author{Juyoung Song}
\affiliation{\mbox{Department of Physics, Konkuk University, Seoul 05029, Republic of Korea}}

\author{Jeyong Park}
\affiliation{Max Planck Institute for Solid State Research, D-70569 Stuttgart, Germany}
\affiliation{School of Natural Sciences, Technische Universit\"at M\"unchen, D-85748 Garching, Germany}

\author{Jinhong Park}
\affiliation{\mbox{Department of Physics, Konkuk University, Seoul 05029, Republic of Korea}}

\date{\today}
\begin{abstract}
We investigate the coexistence of localized states and propagating
one-dimensional (1D) modes in graphene moiré systems. We first show
within a minimal model that a spatially varying scalar potential can
confine localized states, while sign changes of a staggered potential
generate 1D modes along the resulting domain walls. These two types of
states can coexist within the same finite energy window and form a
hybrid network. We then demonstrate a microscopic
realization of this mechanism in twisted monolayer--rhombohedral
$N$-layer graphene. Band structures, energy contours, and Bloch wave functions obtained in a realistic parameter regime reveal
the coexistence of localized nearly flat-band states and propagating
quasi-1D modes. Our results establish twisted
monolayer--rhombohedral graphene as a promising platform for realizing
hybrid electronic networks with coexisting states of distinct
effective dimensionalities.
\end{abstract}
\maketitle
\section{Introduction}\label{sec:Introduction}

Moiré materials formed by stacking van der Waals layers with a small
lattice mismatch or relative twist provide a versatile platform for
engineering electronic band structures and correlated quantum
phases~\cite{Geim2013,Andrei2021}. A prototypical example is
twisted bilayer graphene (TBG), in which nearly flat bands emerge
near the charge-neutrality point at the magic angle
$\theta\simeq1.1^\circ$~\cite{Bistritzer2011,Cao2018a,Cao2018}. The interaction effects enhanced by these
flat bands give rise to a wide variety of correlated and topological
phases~\cite{Yankowitz2019,
Choi2019,Bultinck2020,Xie2021, Song2022}.
More recently, rhombohedral multilayer graphene and
its moiré heterostructures have emerged as a complementary graphene
platform with highly tunable surface flat bands and correlated
topological states, including integer and fractional Chern
insulators~\cite{wang2025fractional,Chen2026LayerEngineered,su2025moire,shen2020correlated,polshyn2020electrical,polshyn2022topological,chen2021electrically, he2021symmetry,liu2020tunable, burg2019correlated, xu2021tunable,lee2019theory,rademaker2020topological,wilhelm2023non,Lu2024FQAH,song2025fractional, Waters2025Chern,Xie2025FCI,phong2025coulomb, wang2025chern,kwan2025textured, park2026tuning,
Zhang2026FlatBand,Li2026HighChern}. In particular, twisted monolayer--rhombohedral $N$-layer graphene
has recently been experimentally realized for several layer numbers
and shown to host a layer-tunable hierarchy of high-Chern-number
orbital and quantum anomalous Hall states~\cite{Liu2025HighChern,Chen2026LayerEngineered,
Wang2026OrbitalMagnetism}.

Besides generating flat bands, moiré structures can spatially
modulate local band gaps and thereby create networks of one-dimensional (1D)
electronic channels. In electrically biased, minimally twisted
bilayer graphene, for example, the $AB$- and $BA$-stacked regions
carry opposite valley Hall indices, and their domain walls support
valley-protected 1D modes~\cite{San-Jose2013,Efimkin2018,Li2016,
Huang2018,DeBeule2021}. The interacting physics of such domain-wall structures has been
investigated using quantum-wire and bosonized network
descriptions~\cite{Hsu2023Network,WangHsu2024}. Related mechanisms
for forming and controlling 1D topological channels have also been
studied in heterostrained and minimally twisted bilayer
graphene~\cite{Georgoulea2024,Hou2024}.
These studies demonstrate that moiré materials can host electronic
structures whose effective dimensionality is lower than that of the
underlying two-dimensional system.

In our previous work~\cite{Park_2023_networkmodel}, we showed that a single
layer of graphene subject to smooth scalar and staggered moiré
potentials can host an unusual coexistence of nearly localized states
and propagating 1D channels within the same finite energy window.
The staggered potential locally opens a gap in graphene and its sign
changes generate valley-resolved chiral modes along the resulting
domain walls. At the same time, the scalar potential confines
additional states near the junctions of the domain-wall network,
giving rise to nearly flat bands. The resulting low-energy degrees
of freedom form a hybrid network consisting of localized modes
connected by propagating 1D channels. In the present work, we extend
this framework to twisted monolayer--rhombohedral $N$-layer graphene and
demonstrate that the coexistence of localized states and propagating
1D modes can also arise in this realistic \moire system.

In this work, we show that twisted monolayer--rhombohedral
$N$-layer 
graphene provides a microscopic realization of this
mechanism. 
We first revisit the minimal monolayer model to clarify how the
spatially modulated scalar and staggered potentials generate,
respectively, localized states and 1D domain-wall modes. We then
derive an effective moiré Hamiltonian for the monolayer graphene by
integrating out the electrically gapped rhombohedral graphene. When
the characteristic decay length of the rhombohedral Green function
is much shorter than the moiré period, the resulting nonlocal
self-energy reduces to a local Hamiltonian of the same form as the
minimal model. From the 
band structures, energy contours, and Bloch wave functions, we demonstrate
that in a realistic
parameter regime, localized nearly flat-band
states coexist with propagating quasi-1D modes, thereby forming a hybrid network. 

Our results establish twisted monolayer--rhombohedral graphene as a
promising platform for realizing a hybrid electronic network with
coexisting localized and propagating degrees of freedom. Such a
system provides a microscopic starting point for interacting network
models in which the 1D channels mediate coupling between localized
spin-, valley-, and orbital degrees of freedom as studied in Ref.~\cite{Park_2023_networkmodel}. 

\section{Toy model: a graphene on a triangle sublattice with twist}
\label{sec:toymodel}

\begin{figure*}[!t]
        \centering
\includegraphics[width=\textwidth]{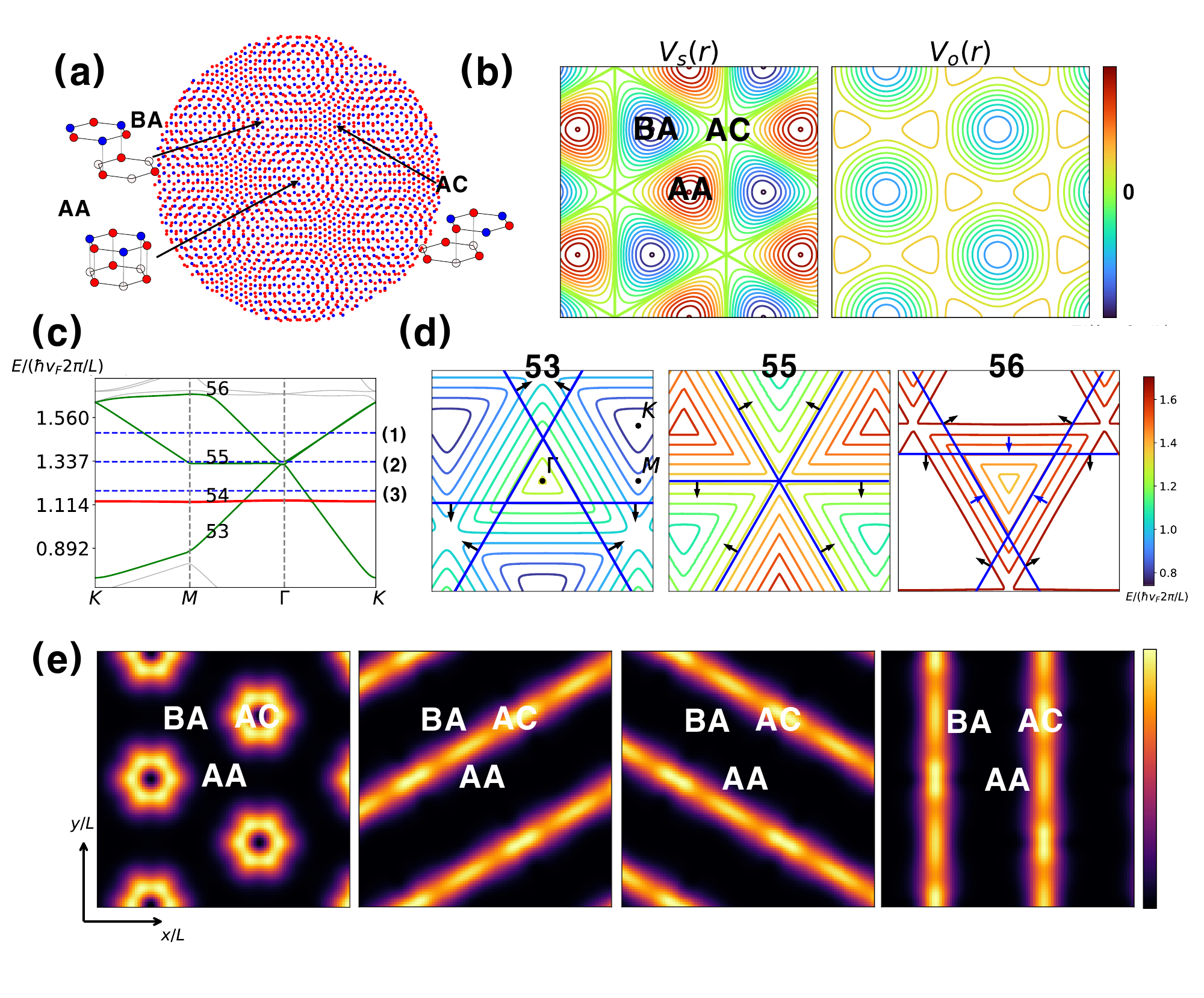}
        \caption{{\bf Potential model}. {\bf (a)} Moir{\'e} pattern from a graphene stacked on a triangular-lattice substrate with the same lattice constant as the graphene. A relative twist between graphene and the substrate produces a spatially varying displacement between the layers, generating local stacking patterns, $BA$, $AC$, $AA$.  
        (b) Spatial profiles of the staggered potential $V_s(\vec{r})$
and scalar potential $V_0(\vec{r})$.
The green lines in the left panel indicate the contours
$V_s(\vec{r})=0$, along which the mass changes sign.
The scalar potential has local minima at the $AC$ regions,
where the domain-wall modes meet.
(c) Low-energy band structure along the high-symmetric path, $K-M-\Gamma-K$ near the $K$ valley.
The green and red curves denote the 1D and flat bands,
respectively. 
(d) Constant-energy contours of the 53, 55, and 56 bands
from left to right.
The blue straight lines represent the ideal, decoupled 1D bands,
and the black arrows indicate the directions in which the
constant-energy contours move with increasing energy. The blue arrows mark the momenta selected for the real-space
probability distributions in (e), while their orientations
indicate the corresponding group-velocity directions. The high-symmetric momentum points, $\Gamma$, $K$, and $M$ are marked in the plot for the 53 band.
(e) Real-space probability distributions of selected Bloch states. The leftmost panel shows the flat-band state at the $K$ point, whereas the three right panels show states in the 56 band at the momenta marked by
    the blue arrows in (d).
    The flat-band state is localized near an $AC$ region, whereas
    the 1D-band states are extended along one direction and confined
    in the perpendicular direction. We choose the parameter $u_0$ as $u_0 \simeq 0.512  \frac{2\pi\hbar v_F}{L}$.
}
        \label{fig1}
    \end{figure*} 
    

We consider a toy model of monolayer graphene placed on a triangular substrate, with a relative twist angle $\theta$ between the graphene and the substrate. The substrate lattice is assumed to have the same lattice constant $a$ as the graphene lattice. The twist produces a spatially varying relative displacement between the two layers, resulting in the moiré pattern illustrated in Fig.~\ref{fig1}(a). The local stacking configurations within a moiré unit cell are labeled by $\alpha\beta$, where the labels $A$ and $B$ (represented by filled red and blue circles, respectively) denote the two graphene sublattices, while $C$ (empty circles) denotes the hollow site at the center of a graphene hexagon. For the triangular substrate, $A$ denotes an atomic site and $C$ denotes a hollow position. The notation $\alpha\beta$ indicates that sublattice $\alpha \in \{A, B, C\}$ of the graphene layer lies directly above sublattice $\beta \in \{A, C\}$ of the substrate. The three local stacking configurations, $AA$, $BA$ and $AC$, are shown in Fig.~\ref{fig1}(a).


For this moir{\'e} configuration, we consider a potential model in which the graphene layer is subject to an electrostatic potential induced by the triangular substrate. We assume that there is no direct hopping or hybridization between the graphene and substrate. Neglecting the spin degree of freedom, the effective low-energy Hamiltonian of graphene in real space is given by
\begin{align} \label{eq:Hamilrealspace}
    H_{0} (\vec{r}) = -iv_F (\sigma_x \tau_z \partial_x  + \sigma_y \tau_0 \partial_y) + V_0 (\vec{r})\mathbbm{1} + V_s(\vec{r}) \sigma_z \tau_0\,.
\end{align}
Here, the Pauli matrices $\tau_i$ and $\sigma_i$ act in the valley and sublattice spaces, respectively, and $v_F$ is the Dirac velocity of graphene. The staggered potential $V_s(\vec{r})$ and scalar potential $V_0(\vec{r})$ are defined as the difference and sum, respectively, of the potentials [$V_{A}(\vec{r})$ and $V_{B}(\vec{r})$]  
acting on the $A$ and $B$ sublattices of graphene: 
\begin{align} \label{eq:staggeredscalarpotential}
    V_s(\vec{r}) \equiv \frac{V_A (\vec{r}) - V_B (\vec{r})}{2}\,, \qquad  V_0 \equiv \frac{V_A (\vec{r}) + V_B (\vec{r})}{2}
\end{align}
The staggered potential $V_s$ locally opens a mass gap in the Dirac spectrum, whereas the scalar potential $V_0$ produces a local energy shift.

At a small twist angle, the local stacking configuration varies smoothly over the moir{\'e} length scale. Consequently, the induced potentials $V_A(\mathbf r)$ and $V_B(\mathbf r)$, and hence $V_0(\mathbf r)$ and $V_s(\mathbf r)$, are also smooth functions of position $\vec{r}$. The smoothness of the potentials allows us to expand $V_\alpha$ with the lowest Fourier components: 
\begin{align}
V_{\alpha} (\vec{r})= \overline{V}_{\alpha} +  \sum_{\beta = A, C} \sum_{i=1}^6 u_{\alpha \beta} e^{i \vec{G}_i \cdot (\vec{r}-\vec{r}_{\alpha \beta})}\,,
\end{align}
with $\alpha\in \{A, B\}$. 
Here $\vec{G}_i$ $(i=1, \cdots, 6)$ are the six shortest nonzero reciprocal lattice vectors of the moir{\'e} superlattice with $|\vec{G}_i| = G = \frac{4\pi}{\sqrt{3}L}$ and $L$ is the moir{\'e} lattice constant. 
The vectors $\vec{r}_{\alpha \beta}$ denote the positions of the corresponding local stacking configurations: $\vec{r}_{AA} = \vec{r}_{BC} = \vec{0}$ and $\vec{r}_{AC} = - \vec{r}_{BA} = (\frac{\sqrt{3}L}{6}, \frac{L}{2})^T$.
The parameters $u_{\alpha \beta}$ represent the amplitudes of the first moir{\'e} harmonics associated with the local stacking configurations $\alpha \beta$.

As a minimal parametrization, we neglect lattice-relaxation-induced corrugation and assume
\begin{align}
u_{AA}=u_{BA}=u_0,
\qquad
u_{AC}=u_{BC}=0.
\end{align}
We also neglect the spatially uniform components $\overline{V}_{\alpha}$ by assuming that the $A$ and $B$ sublattices of graphene experience the same potential on average over a 
\moire unit cell. Under this assumption, the uniform staggered potential $\overline{V}_{s} \equiv (\overline{V}_A - \overline{V}_B)/2$ vanishes, while the remaining uniform scalar potential $\overline{V}_{0} \equiv (\overline{V}_A + \overline{V}_B)/2$ can be absorbed into the chemical potential. However, when we consider a more realistic system in Sec.~\ref{sec:twistedmonolayerbilayer}, we will show that $\overline{V}_s$ becomes finite and has to be included explicitly. 

The staggered and scalar potentials [Eq.~\eqref{eq:staggeredscalarpotential}] are plotted in Fig.~\ref{fig1}(b). The staggered potential $V_s (\vec{r})$ shown in the left panel of Fig.~\ref{fig1}(b) vanishes along the green lines and changes sign across them. According to the Jackiw--Rebbi mechanism~\cite{Jackiew1976}, each mass domain wall supports a 1D chiral gapless mode for each valley. The modes originating from the $K$ and $K'$ valleys propagate in opposite directions, forming a pair of counter-propagating valley-polarized modes. For a \moire system with a large \moire period $L$, the potentials vary smoothly on the \moire scale and therefore strongly suppress inter-valley scattering. Consequently, the two spatially overlapping 1D modes remain approximately decoupled.

At the same time, the scalar potential $V_0 (\vec{r})$ (the right panel) has a minimum value at the $AC$ region where the 1D domain-wall modes meet. The quadratic scalar potential near the $AC$ region confines electrons, which may lead to localized modes at the junction.

These observations on the staggered and scalar potentials suggest an effective network model in which propagating 1D domain-wall modes coexist and couple with localized modes at the junctions. The emergence of such a network model was discussed in Ref.~\cite{Park_2023_networkmodel}. The present network model shares the same structure as that of Ref.~\cite{Park_2023_networkmodel}, but it is shifted by $\vec{r} \rightarrow \vec{r} + \vec{r}_{AC}$ compared with that of Ref.~\cite{Park_2023_networkmodel}.


The low-energy band structure of this toy model can be obtained using the approach first developed in Ref.~\cite{Bistritzer2011}. 
To this end, we Fourier transform Hamiltonian~\eqref{eq:Hamilrealspace} to the momentum space and
diagonalize the resulting Hamiltonian. The band structure in the $K$ valley is plotted along the
high-symmetry path $K$--$M$--$\Gamma$--$K$ of the monolayer graphene in
Fig.~\ref{fig1}(c). The bands fall into two classes: 1D bands (denoted as green lines) and flat bands (red lines). We now investigate these bands by inspecting the constant-energy contour of the bands and the Bloch wave function.

The constant-energy contours of the three 1D bands are shown
in Fig.~\ref{fig1}(d): 53, 55, and 56th bands (green lines in Fig.~\ref{fig1}(c)) are displayed from the left to the right panels of Fig.~\ref{fig1}(d). The resulting contours have approximately equilateral
triangular shapes centered at either the $K$ or $\Gamma$ point.
The bands can be understood as 1D bands weakly coupled to each other. To see the 1D nature, we consider three ideal, decoupled 1D bands
related by the $C_{3z}$ rotational symmetry, as shown as the three green straight lines in each panel of Fig.~\ref{fig1}(d).
The three contours are evaluated at the energies indicated by
the dashed lines labeled as (1), (2), and (3) in
Fig.~\ref{fig1}(c), respectively.
In the decoupled 1D limit, the energy is independent of the
momentum parallel to each straight line and disperses
only in the perpendicular direction, as indicated by the blue
arrows. When a small coupling between them is induced, the bands hybridize near the three intersection points and
develop avoided crossings, causing the constant-energy contours
to separate and become rounded near these points. As shown in the constant-energy contours, 
such 1D bands with a small coupling emerge in our potential model. At the same time, the 54th band, shown in red, is nearly flat. The spectrum in the $K'$ valley (not shown) is related to that in the
$K$ valley by time-reversal symmetry. 

The features of these exotic bands can also be understood by inspecting the real-space probability distribution $P_{\vec{k}} (\vec{r}) = \sum_{\alpha = A,B}|\psi_{\vec{k}, \alpha}|^2 (\vec{r})$  of Bloch wave functions, as shown in Fig.~\ref{fig1}(e). The leftmost panel of Fig.~\ref{fig1}(e)
represents $P_{\vec{k}} (\vec{r})$ for the flat band (54 band) at $K$ point. As expected from the effective model, the wave function is highly localized in the $AC$ region. The remaining three panels of Fig.~\ref{fig1}(e) show
$P_{\vec{k}}(\vec{r})$ for the 56th band at the three momenta
marked by the blue arrows in Fig.~\ref{fig1}(d): The orientation
of each arrow indicates the group velocity at the corresponding
momentum. The wave functions are extended along the corresponding
group-velocity directions and confined in the perpendicular
directions.

\begin{figure}[h]
    \centering
\includegraphics[width=0.3\textwidth]{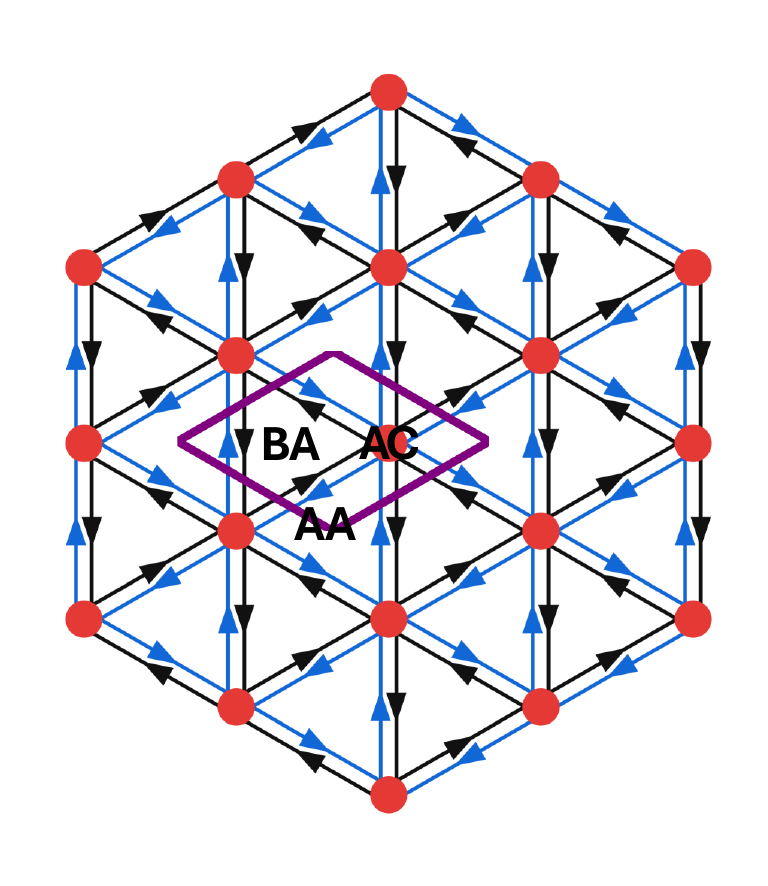}
    \caption{{\bf Network model} emerging from the potential model presented in Fig.~\ref{fig1}. Localized modes (denoted red nodes) in the $AC$ region are connected by pairs of counterpropagating 1D modes. The blue and black arrows represent the chiral modes originating from the $K$ and $K'$ valleys, respectively, and indicate their propagation directions.}
    \label{fig:networkmodel}
\end{figure}

From the above discussions of the band structure,
constant-energy contours, and Bloch wave functions,
we conclude that an effective network model consisting of
1D domain-wall modes and localized modes emerges within a
finite energy window; see figure~\ref{fig:networkmodel}.
The nearly flat band is associated with states localized near
the $AC$ junctions, while the links of the network are formed
by pairs of counterpropagating 1D modes originating from the
$K$ and $K'$ valleys. The $K$-valley modes are explicitly demonstrated in
Fig.~\ref{fig1}, while the corresponding $K'$-valley modes
follow from time-reversal symmetry. Consequently, the 1D modes
in the two valleys propagate in opposite directions along each
domain wall. 





\section{Twisted monolayer--rhombohedral $N$-layer graphene}
\label{sec:twistedmonolayerbilayer}

So far, we have discussed a toy model in which an effective
network consisting of 1D modes and localized modes emerges.
As a possible platform for realizing this model, we consider
twisted monolayer--rhombohedral $N$-layer graphene, which has
recently been experimentally realized for several layer
numbers~\cite{Liu2025HighChern,Chen2026LayerEngineered,
Wang2026OrbitalMagnetism}. In this system, monolayer graphene
is placed on rhombohedral $N$-layer graphene with a relative
twist angle. Compared with the toy model, the triangular
substrate is replaced by rhombohedral graphene. A perpendicular displacement field is applied to open a gap
in the low-energy spectrum of the rhombohedral graphene. When
the energy lies well inside this gap, a real charge transfer between the monolayer and rhombohedral
graphene is strongly suppressed. In this regime, the effect
of the rhombohedral graphene on the monolayer can be described
by an effective \moire potential, analogous to the
potential model introduced in Sec.~\ref{sec:toymodel}.

\begin{figure*}[t]
        \centering
\includegraphics[width=\textwidth]{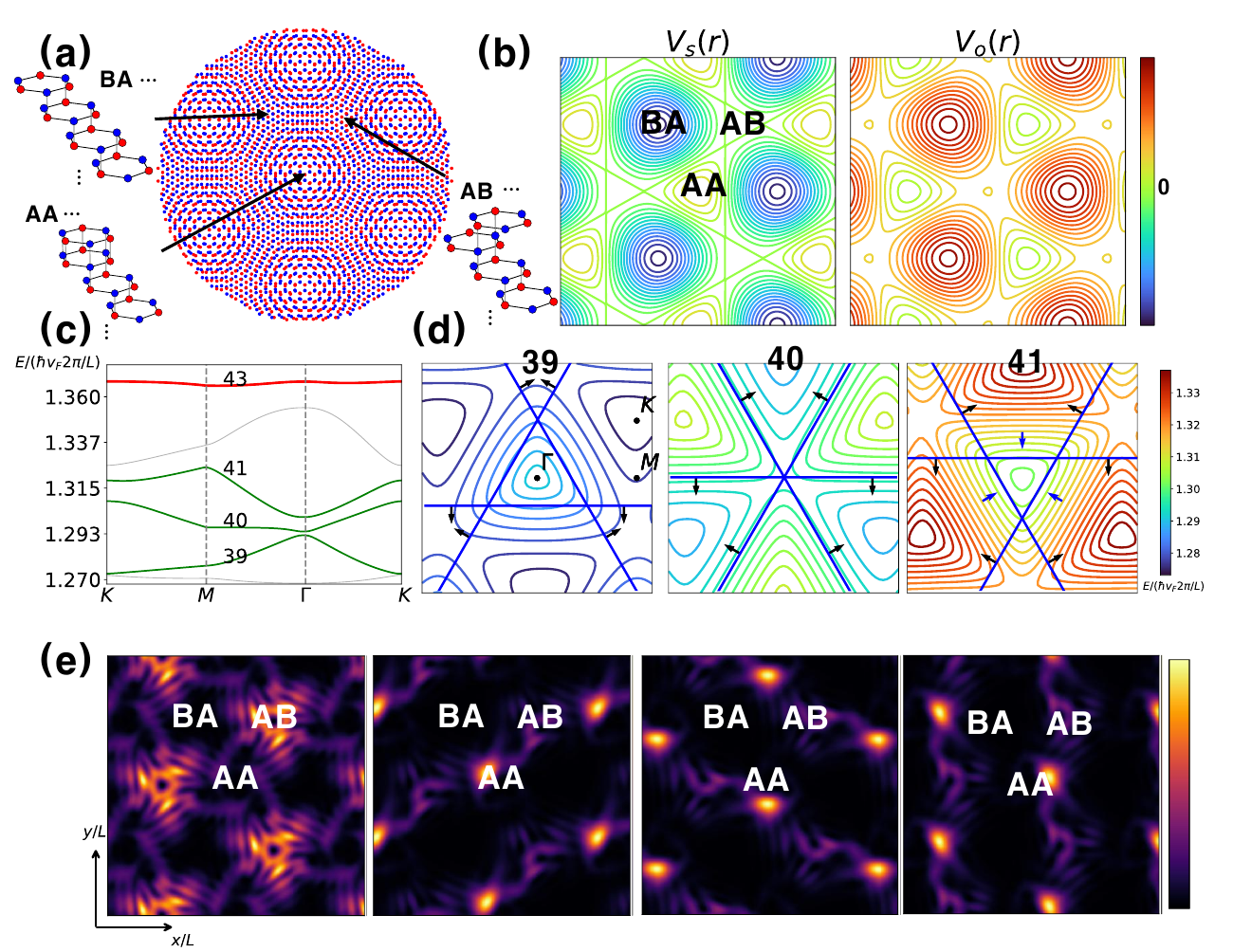}
        \caption{{\bf Twisted rhombohedral--monolayer graphene}. \textbf{(a)} Schematic atomic structure of monolayer graphene on top of rhombohedral $N$-layer graphene with a relative twist angle. Red and blue circles denote
the two sublattices, respectively. The enlarged side-views show the representative local
$AA$, $AB$, and $BA$ stacking configurations.
\textbf{(b--e)} Results for $N=3$.
\textbf{(b)} Contour plots of the effective staggered potential
$V_s(\vec{r})$ and scalar potential $V_0(\vec{r})$ induced in the
monolayer graphene. The light-green contours 
indicate the
domain walls defined by $V_s(\vec{r})=0$, which separate regions
with opposite signs of the staggered potential. 
\textbf{(c)} Moir{\'e} band structure along the high-symmetry momentum path
$K$--$M$--$\Gamma$--$K$. 
The green and red bands identify the
1D and nearly flat bands, respectively. The four bands
in the energy window of interest are labeled by their band indices
$39$--$41$ and $43$.
\textbf{(d)} Constant-energy contours of bands $39$, $40$, and $41$ in the
\moire Brillouin zone. The blue lines show the corresponding contours
of an idealized 1D dispersion, providing a reference
for the degree to which each band approaches the ideal 1D
limit. The arrows indicate the directions of the group velocities,
and the color scale represents the band energy.
\textbf{(e)} Real-space probability densities of representative eigenstates.
The leftmost panel shows an eigenstate of the nearly flat band $43$ at the $K$ point. The remaining three panels show eigenstates selected at
momenta on the blue arrows in \textbf{(d)}. The parameters are $v_F = 8.45\times10^5 \ \text{m/s}, \ v_3 = 9.16\times 10^4 \ \text{m/s}, \ v_4 = 4.46 \times 10^4 \ \text{m/s},$ $U=-200~\mathrm{meV}$,
$\gamma=361~\mathrm{meV}$, $w_{AA}=55~\mathrm{meV}$, \ 
$w_{AB}=110~\mathrm{meV}$, $\Delta' = 0 \ \text{meV}$, and $\theta = 0.181^\circ$. The values of the parameters $\gamma$, $v_3$, and $v_4$ are extracted from Ref.~\cite{Jung2014}. Energies in \textbf{(c)} and \textbf{(d)} are expressed
in units of $ 2\pi\hbar v_F/L$.} \label{fig:rhomobohedralmain}
    \end{figure*}

The rhombohedral $N$-layer graphene consists of $N$ 
layers of graphene with the stacking sequence $ACBA\cdots$. As in the toy
model, we list the stacking sequence from the uppermost to the
lowermost layer and adopt the anticyclic stacking convention
for rhombohedral graphene. When monolayer graphene is stacked
on the rhombohedral $N$-layer graphene with a relative twist angle $\theta$,
three high-symmetry local stacking configurations appear within
a \moire unit cell: $AAC\cdots$, $BAC\cdots$, and $ABA\cdots$,
as shown in Fig.~\ref{fig:rhomobohedralmain}(a). For brevity, we refer to
these configurations as the $AA$, $BA$, and $AB$ regions,
respectively.

To capture the essential low-energy physics, we employ a
continuum model following the approach developed in
Ref.~\cite{Bistritzer2011}. We consider an electric field
applied perpendicular to the $N+1$ graphene layers. Assuming
a linear potential drop across the layers, the electrostatic
potential on the $i$th layer, counted from the uppermost layer,
is taken to be 
\begin{align}
    U_i = U \Big[\frac{1}{2} - \frac{i-1}{N} \Big]\,, \qquad
    1\leq i\leq N+1.
\end{align}
Here $U$ is the potential difference between the uppermost and lowermost layers. The total Hamiltonian in a continuum limit reads
\begin{align} \label{eq:Hamiltoniantmrhg}
   H= \begin{pmatrix}
H_{\mathrm{MLG}} & T \\
T^\dagger & H_{\mathrm{RHG}}
\end{pmatrix}\,,
\end{align}
where $H_{\mathrm{MLG}}$ and $H_{\mathrm{RHG}}$ describe the
monolayer graphene and the rhombohedral $N$-layer graphene,
respectively, while $T$ denotes the \moire
tunneling between them. Near the $K$ valley, the monolayer
Hamiltonian $H_{\mathrm{MLG}}$ in the sublattice basis is given by
\begin{align} \label{eq:monolayerHamiltonian}
    H_{\mathrm{MLG}} (\vec{k}) = v_F\vec{k}_{\theta/2}\cdot \vec{\sigma} + \frac{U}{2} \mathbbm{1}\,, 
\end{align}
with the rotated momenta $\vec{k}_\theta \equiv R(\theta)\vec{k}$ by twist angle $\theta$. Here $R(\phi)$ is the rotation matrix with angle $\phi$, and
$\vec{\sigma}=(\sigma_x,\sigma_y)$ are the Pauli matrices acting on the sublattice space. The  rhombohedral $N$-layer graphene is described in the basis
$(A_1,B_1,\ldots,A_N,B_N)$, where the indices
$1,\ldots,N$ label the layers from top to bottom, by the
following $2N\times2N$ Hamiltonian:
\begin{align} \label{eq:rhombohedralham}
H_{\mathrm{RHG}}(\vec{k})
=
\begin{pmatrix}
h_1(\vec{k}) & \Gamma & 0 & \cdots & 0 \\
\Gamma^\dagger & h_2(\vec{k}) & \Gamma & \ddots & \vdots \\
0 & \Gamma^\dagger & h_3(\vec{k}) & \ddots & 0 \\
\vdots & \ddots & \ddots & \ddots & \Gamma \\
0 & \cdots &0 & \Gamma^\dagger & h_N(\vec{k})
\end{pmatrix},
\end{align}
where the intralayer Hamiltonian of the $j$th layer is
\begin{align}
h_j(\vec{k})
=
v_F\vec{k}_{-\theta/2}\cdot\vec{\sigma}
+
U_{j+1}\mathbbm{1} 
\,.
\end{align}
Since the rhombohedral block starts from the second layer of
the full heterostructure, its $j$th layer experiences the
potential $U_{j+1}$.
For the anticyclic stacking sequence $ACBA\cdots$, the
nearest-neighbor 
interlayer hopping is given by 
\begin{align}
    \Gamma = \begin{pmatrix}
        0 & 0 \\ \gamma & 0
    \end{pmatrix}\,.
\end{align}
Here, $\gamma$ 
denotes the nearest-neighbor 
interlayer hopping
between $A_{j+1}$ and $B_{j}$.
In this minimal model, we neglect remote hopping processes 
and the onsite-energy asymmetry $\Delta'$; see Ref.~\cite{Jung2014} for the neglected parameters. 
However, we have checked that the inclusion of the neglected parameters does not qualitatively change the physics we discussed in the paper. 
The \moire tunneling between the monolayer graphene and the
uppermost layer of the rhombohedral graphene is described by the $2 \times (2N)$ matrix 
\begin{align} \label{eq:tunnelingoperator}
    T (\vec{r}) = \sum_{m=1}^3 e^{i \vec{q}_m \cdot \vec{r}} T_m \,,
\end{align}
with 
\begin{align}
    T_m = \begin{pmatrix}
        w_{AA} & w_{AB} e^{-i (m-1)\frac{2\pi}{3}} & 0 & 0 &\cdots & 0 & 0 \\
         w_{AB}  e^{i (m-1)\frac{2\pi}{3}} & w_{AA} & 0 & 0 &\cdots & 0 & 0 
    \end{pmatrix}\,,
\end{align}
with $m = 1, 2,3$. Here, $w_{AA}$ denotes the tunneling strength in the local $AA$ stacking configuration, while $w_{AB}$ denotes the tunneling strength in the $AB$ and $BA$ configurations. We retain the three dominant momentum-transfer processes~\cite{Bistritzer2011}. The vectors $\vec{q}_m$ denote the three momentum transfers corresponding to the dominant interlayer tunneling processes; $\vec{q}_1 = q_{\theta} (0, 1)$, $\vec{q}_2 = q_{\theta} (-\frac{\sqrt{3}}{2}, -\frac{1}{2})$, $\vec{q}_3 = q_{\theta} (\frac{\sqrt{3}}{2}, -\frac{1}{2})$, and $q_{\theta} = \frac{8\pi \sin(\theta/2)}{3a} = \frac{4\pi}{3L}$, $a$ is the lattice constant, and $L = \frac{a}{2\sin(\theta/2)}$ is the \moire lattice constant.  


We now map the Hamiltonian \eqref{eq:Hamiltoniantmrhg} for the twisted monolayer--rhombohedral graphene into the effective Hamiltonian \eqref{eq:Hamilrealspace} for the graphene by using the second order perturbation theory (namely, the Schrieffer-Wolff transformation).
When energy is inside the gap of the rhombohedral graphene, i.e., $U_{N+1} < E <U_2$ and $w_{AA}, w_{AB} \ll |U_{2}-U_{N+1}|$, the direct hybridization with the rhombohedral graphene is not allowed. Then, the leading correction to the monolayer Hamiltonian, \eqref{eq:monolayerHamiltonian}, comes from the second-order process in tunneling, resulting in the effective Hamiltonian for the monolayer graphene,
\begin{subequations}
\label{eq:effectiveHammomentum}
\begin{align} 
(H_{\text{eff}})_{\vec{k}\vec{k'}} &\approx H_{\text{MLG}} \delta_{\vec{k}\vec{k}'}  + (\Delta H_{\text{MLG}})_{\vec{k}\vec{k}'}\,, \\ 
    (\Delta H_{\text{MLG}})_{\vec{k}\vec{k}'}&=\sum_{m,m'=1,2,3}T_m\frac{1}{E-H_{\text{RHG}}} T_{m'}^\dagger \delta_{\vec{k'},\vec{k}+\vec{q}_m - \vec{q}_{m'}}\,. 
\label{eq:effectiveHammomentumcorrection}
\end{align}
\end{subequations}
Transforming into the real space, the second term of Eq.~\eqref{eq:effectiveHammomentum} becomes 
\begin{align}
    \Delta H_{\text{MLG}} (\vec{r}, \vec{r}') = T(\vec{r}) g_{\text{RHG}}(E; \vec{r}-\vec{r}') T^\dagger (\vec{r}')\,,
\end{align}
where $g_{\text{RHG}}(E; \vec{r}-\vec{r}') = (E- H_{\text{RHG}})^{-1} (\vec{r}-\vec{r}')$ is the bare Green function of the rhombohedral graphene in real space. 

Since the energy $E$ lies inside the gap of the rhombohedral
graphene, the Green function $g_{\mathrm{RHG}}
(E;\vec{r}-\vec{r}') \sim e^{-|\vec{r}-\vec{r}'| /\xi}$ decays at large $|\vec{r}-\vec{r}'|$. Its decay length is given by
$\xi \equiv 1/|\operatorname{Im}[\vec{k}_0]|$, where
$\vec{k}_0$ is the complex-momentum solution with the smallest
$|\text{Im}[\vec{k}]|$ among those satisfying
\begin{align}
    \det\left[E-H_{\mathrm{RHG}}(\vec{k})\right]=0.
    \label{eq:determinant}
\end{align}
Solving Eq.~\eqref{eq:determinant} with the rhombohedral graphene Hamiltonian \eqref{eq:rhombohedralham}, 
we find that in the regime of $U \ll \gamma$, $\xi$ scales as 
\begin{align}
    \xi
    \sim
    \frac{v_F}{
    \gamma^{1-\frac{1}{N}}
    \left[
        (U_2-E)(E-U_{N+1})
    \right]^{\frac{1}{2N}}
    }\sim \frac{v_F}{\gamma (U/\gamma)^{1/N}}\,.
\end{align}
For sufficiently small twist angles, $\xi$ is much smaller than $L$, and thus the nonlocal effect of $\Delta H_{\text{MLG}} (\vec{r}, \vec{r}')$ can be neglected. We also note that, for $U\ll\gamma$, $\xi$ decreases as $N$ increases. The local approximation improves with increasing $N$.


In this local approximation, we can focus on $\Delta H_{\text{MLG}} (\vec{r}, \vec{r})$ and map it onto the effective Hamiltonian as 
\begin{align} \label{eq:effectiveHamTMRHG}
 H_{\text{eff}} (\vec{r}) = v_F(-i\vec{\nabla}_{\vec{r}} - e\vec{A}(r)) \cdot \vec{\sigma}  + V_0(\vec{r}) \mathbbm{1} +V_s(\vec{r}) \sigma_z\,.
\end{align}
Here, we approximated $V_0(\vec{r}) \simeq \frac{1}{2} \text{Tr}[H_{\text{MLG}} (\vec{r}, \vec{r}) \mathbbm ] $, $V_s (\vec{r}) \simeq \frac{1}{2} \text{Tr}[H_{\text{MLG}} (\vec{r}, \vec{r}) \sigma_z]$, and the effective vector potentials $\vec{A} = (A_x, A_y)$ where
$A_l \simeq \frac{1}{2} \text{Tr}[H_{\text{MLG}} (\vec{r}, \vec{r}) \sigma_l]$ with $l = x,y$. We assumed that the twist angle $\theta$ is small and therefore neglect the
$\theta$-dependence in the first term of Eq. \eqref{eq:effectiveHamTMRHG}. We have checked that the effective magnetic field  $\vec{B} = \vec{\nabla}_{\vec{r}} \times \vec{A}$ from the obtained vector potentials vanishes in every $\vec{r}$, and thus we can gauge away the vector potentials. Evaluating the scalar potential $V_0$ and staggered potential $V_s$ from Eqs.~\eqref{eq:tunnelingoperator}, \eqref{eq:effectiveHammomentum} and using Eq.~\eqref{eq:staggeredscalarpotential}, we show that $V_A$ and $V_B$ take the form as 
\begin{subequations}
\label{eq:potentialrhombohedral}
\begin{align}
    V_A(\vec{r}) &= \overline{V}_0+\overline{V}_s + \sum_{\alpha = A,B}\sum_{i=1}^6 u_{A \alpha}e^{i\vec{G}_i\cdot (\vec{r}-\vec{r}_{A\alpha})}\,, \\ 
V_B(\vec{r}) &= \overline{V}_0-\overline{V}_s+ \sum_{\alpha = A,B}\sum_{i=1}^6 u_{B \alpha}e^{i\vec{G}_i\cdot (\vec{r}-\vec{r}_{B\beta})},
\end{align}
\end{subequations}
where we write the parameters $u_{\alpha \beta}$, $\overline{V}_0$ and $\overline{V}_s$,  
\begin{subequations}
\begin{align}
    &u_{AA} = \frac{2N \omega^2_{AA}}{U(1-N)}, \quad u_{BA} = \frac{2N \omega^2_{AB}}{U(1-N)}
    \\ &u_{AB} = -\frac{2\omega^2_{AB}(N-3)NU}{\zeta}\,,  \\ &
    u_{BB} = -\frac{2\omega^2_{AA}(N-3)NU}{\zeta} \label{eq:effectiveparameters}
\\&\overline{V}_s = \frac{3(\omega^2_{AA}-\omega^2_{AB})(8N^3\gamma^2)}{2U(N-1)\zeta}
    \\&\overline{V}_0 = \frac{3(\omega^2_{AB}+\omega^2_{AA})(8N^3\gamma^2-4(N-3)(N-1)NU^2)}{2U(N-1)\zeta} 
\end{align}
\end{subequations}
with $\zeta = [(N-3)(N-1)U^2-4N^2\gamma^2]$. Here $E$ is set to the middle of the gap, i.e., $E = (U_2 + U_{N+1})/2$. 

In contrast to the toy model discussed in Sec.~\ref{sec:toymodel}, the uniform staggered potential is generally finite in the presence
of lattice relaxation with $w_{AA}\neq w_{AB}$, owing to the
inequivalence of the two surface sublattices of rhombohedral graphene. The terms with $m=m'$ in Eq.~\eqref{eq:effectiveHammomentumcorrection} involve no momentum transfer within the
monolayer graphene and therefore give the spatially uniform
part of the effective potential. Its staggered component is
given by
\begin{align}
    \overline{V}_s(\vec{k})
    =
    \frac{1}{2}
    \operatorname{Tr}
    \left[
        \sigma_z
        (\Delta H _{\text{MLG}})_{\vec{k}\vec{k}}
    \right]\,. 
\end{align}
Evaluating this expression shows that a finite uniform
staggered potential requires two ingredients simultaneously: (i)
the inequivalence of the two surface sublattices of the
rhombohedral graphene and (ii) the tunneling anisotropy
$w_{AA}\neq w_{AB}$. If $w_{AA}=w_{AB}$, the two monolayer
sublattices couple to the inequivalent substrate channels with
equal total weights, and their uniform energy shifts cancel to each other. Conversely, if sublattice symmetry were present in the rhombohedral graphene, the $AB$ and $BA$ configurations would be equivalent, and the $A$ and $B$ sublattices of graphene experience the same potential on average over a \moire unit cell even when
$w_{AA}\neq w_{AB}$. Thus, the uniform staggered potential
arises only from the combined effect of the substrate
sublattice asymmetry and the tunneling anisotropy.

We now demonstrate that the coexistence of 1D and
localized states found in the toy model also appears in a realistic
twisted rhombohedral--monolayer graphene system. For concreteness, we
consider $N=3$ with $v_F = 8.45\times10^5 \ \text{m/s}, \ v_3 = 9.16\times 10^4 \ \text{m/s}, \ v_4 = 4.47 \times 10^4 \ \text{m/s},$ $U=-200\mathrm{meV}$,
$\gamma=361\mathrm{meV}$, $w_{AA}=55\mathrm{meV}$,
$w_{AB}=110\mathrm{meV}$, $\Delta' = 0 \ \text{meV}$, and $\theta = 0.181^\circ$. We use the values of the parameters $\gamma$, $v_3$, $v_4$ extracted from Ref.~\cite{Jung2014}. 
In our notation, $\gamma$ corresponds to $t_1$ in Eq.~(26) of Ref.~\cite{Jung2014}, while $v_3$ and $v_4$ are the velocities associated with the remote hopping parameters $t_3$ and $t_4$, respectively.
We set $\Delta' = 0$ in the present calculation, but we have checked that the inclusion of a finite $\Delta'$ does not qualitatively change the results. Our choice of $w_{AA}/w_{AB}=0.5$ is motivated by the suppression of the effective
$AA$ tunneling relative to the $AB/BA$ tunneling found in relaxed small-angle graphene interfaces. ~\cite{Carr2019Exact,Walet2019Channels,rademaker2020topological}. 

Figure~\ref{fig:rhomobohedralmain}(b) shows the
resulting staggered and scalar potentials obtained from Eq.~\eqref{eq:potentialrhombohedral}.
Lattice relaxation generates a substantial uniform component
$\overline{V}_s$, thereby shifting the zero contours of the total
staggered potential $V_s(\vec r)$. Nevertheless, the spatially varying components of the staggered potential  
remains sufficiently strong to produce sign changes near the $AB$ region. The contours
$V_s(\vec r)=0$, shown in light-green lines, therefore form a network of domain
walls along which 1D modes can emerge. At the same
time, the scalar potential $V_0(\vec r)$ develops extrema near the
$AB$ regions, suggesting the formation of localized modes there. We note that the rhombohedral gap, $\Delta_{\rm rh}\sim 67$ meV, is smaller than $w_{AA}=110$ meV. Thus, the system is formally outside the strict perturbative regime. Nevertheless, the numerically obtained band structure shown below is well captured by the effective model in Eq.~\eqref{eq:potentialrhombohedral}.


The corresponding \moire band structure is shown in
Fig.~\ref{fig:rhomobohedralmain}(c). The bands near the energy window of interest
contain both dispersive 1D bands, highlighted in
green, and a nearly flat band, highlighted in red. 
To reveal their
dimensional characteristics more directly, we plot in
Fig.~\ref{fig:rhomobohedralmain}(d) the energy contours of bands $39$--$41$.
For an ideal 1D dispersion, the constant-energy contours form straight lines, shown in blue. Some of the calculated contours develop
extended segments that follow these lines, demonstrating the
emergence of quasi-1D propagation along the domain-wall
network. This correspondence is particularly clear for band $40$ and $41$,
whose contours closely approach those of the ideal 1D
dispersion. 

The real-space probability densities in
Fig.~\ref{fig:rhomobohedralmain}(e) further confirm this interpretation. The
leftmost panel shows a representative nearly flat-band state at the
$K$ point, whose probability density is concentrated near the $AB$
regions. The remaining panels show states taken at three
$C_{3z}$-related momenta on the 1D bands indicated by
the blue arrows in Fig.~\ref{fig:rhomobohedralmain}(d). Their probability densities are concentrated along three directions of the domain-wall network, which intersects near the $AB$ region. 
The enhanced probability weight near the junctions can be attributed
to the reduced propagation velocity caused by strong hybridization
among the intersecting 1D modes. These results demonstrate that the realistic
twisted multilayer graphene system retains the characteristic coexistence of localized
states and propagating 1D modes anticipated from the toy
model.

\section{Conclusion and outlook}
In conclusion, we have shown that a van der Waals \moire
heterostructure can host a hybrid electronic network in which localized states and
quasi-1D modes coexist within a finite energy window.
Within a minimal toy model, the scalar potential confines localized
states, whereas sign changes of the staggered potential generate
one-dimensional modes along the resulting domain walls. We further
demonstrated that the same mechanism is realized in twisted
monolayer--rhombohedral graphene. By integrating out the electrically gapped
rhombohedral graphene, we derived an effective \moire Hamiltonian
acting on the monolayer graphene, which reduces to the
toy-model Hamiltonian. The calculated
band dispersions, constant-energy contours, and real-space
probability densities confirm the coexistence of localized states
and propagating quasi-1D modes, establishing twisted
monolayer--rhombohedral graphene as a realistic platform for
realizing a network of electronic states with distinct effective
dimensionalities.

The coexistence of localized states and propagating 1D channels also
provides a promising setting for exploring interaction-driven
phenomena beyond the single-particle physics studied here. When the
coupling between the two sectors is weak, the 1D channels can mediate
interactions between the localized spin and valley degrees of
freedom, potentially stabilizing unconventional spin--valley orders,
including states with nonzero scalar spin chirality, as discussed in
Ref.~\cite{Park_2023_networkmodel}. At stronger coupling, the system
may instead enter a Kondo-lattice regime in which the localized
moments are dynamically screened by the 1D electronic modes. Since
the latter are generically described as Luttinger liquids rather than
Fermi liquids, their coupling to the localized moments may lead to
unconventional quantum critical behavior and non-Fermi-liquid
physics. Magnetic fluctuations associated with the localized degrees of
freedom may further influence pairing tendencies in the propagating
1D channels, providing a possible route toward unconventional
superconductivity. The hybrid network identified here
therefore offers a versatile platform for investigating the interaction effects arising from the coupling between localized degrees of freedom and topological 1D
modes.

\section*{Acknowledgements}
J.Y.S and J.H.P acknowledge support from Korea NRF (Grant No. RS-2026-25492880).
J.Y.P thanks the Kwanjeong Educational Foundation for support.



\bibliography{Refs.bib}

\end{document}